\documentstyle[prb,aps,epsfig,twocolumn]{revtex}

\begin{document}
\title{Lattice instabilities of PbZrO$_3$/PbTiO$_3$ [1:1]
superlattices from first principles}

\author{Claudia Bungaro and K. M. Rabe}
\address
{Department of Physics and Astronomy, Rutgers University, 
Piscataway, NJ 08854-8019}

\date{\today}
\maketitle
\begin{abstract}

Ab initio phonon calculations for the 
nonpolar reference structures of the (001), (110), and (111)
PbZrO$_{3}$/PbTiO$_{3}$ [1:1] superlattices are presented.  The
unstable polar modes in the tetragonal (001) and (110) structures are
confined in either the Ti- or the Zr-centered layers and display
two-mode behavior, while in the cubic (111) case one-mode behavior is
observed.  Instabilities with pure oxygen character are observed in
all three structures.  The implications for the ferroelectric behavior
and related properties are discussed.

\end{abstract}


\section{Introduction}

The soft-mode theory of ferroelectrics \cite{Lines} has 
established the relationship of the unstable polar modes of a
high-symmetry nonpolar reference structure to the physics of both
the ferroelectric transition and the large dielectric and
piezoelectric coefficients.  
First-principles density functional perturbation theory (DFPT),
\cite{Zein,BGT} permits the direct quantitative investigation of such
unstable modes and the accurate computation of the
interatomic force constants, phonon frequencies and dynamical matrix
eigenvectors of the cubic reference structures of ferroelectric
perovskite oxides such as BaTiO$_{3}$, PbTiO$_{3}$, SrTiO$_{3}$, and
KNbO$_{3}$, has lead to a deeper understanding of the individual compounds
and materials trends\cite{GhosezBT,Comparative,KrakauerST,KrakauerKN}.

For technological applications requiring high-performance dielectrics,
piezoelectrics, and ferroelectrics, the most promising candidates are
not pure perovskite compounds, but solid solutions, such as
Pb(Zr$_{x}$Ti$_{1-x}$)O$_3$ (PZT)\cite{PZT},
(Ba$_{x}$Sr$_{1-x}$)TiO$_3$ (BST)\cite{BST} and PMN-PT\cite{PMNPT},
and atomic-scale superlattices including BaTiO$_{3}$/SrTiO$_{3}$ and
KNbO$_{3}$/KTaO$_{3}$.\cite{superlattices} In these systems, the
unstable polar modes of their high-symmetry nonpolar reference structure
are similarly expected to be central to a predictive understanding of
their properties.  However, their chemical and structural complexity,
requiring the use of large supercells, leads to a tremendous increase
in the computational demands of first-principles calculations.

Accurate first-principles calculations can be performed only
for short period superlattices and the simplest ordered configurations
of solid solutions.  PZT is a natural choice for first-principles
investigation.  Not only is it already in widespread use in
piezoelectric transducer applications, but its optimal composition, at
the morphotropic phase boundary, is almost exactly x=0.5, which
corresponds to ordered structures with supercells as small as double
the primitive perovskite unit cell.  Indeed, work on the (001) and
(111) doubled supercells dominates available first-principles results
on PZT.\cite{Cohen,Bellaiche1,Bellaiche2,Rappe,Fornari} While the
atomic arrangements in real bulk PZT are of course much more
disordered, the study of the lattice dynamics of ordered structures
with [1:1] compositional modulation along various crystallographic
directions provides (i) an accurate characterization of the phonons of
ultrathin superlattices, (ii) a guide to understanding the vibrational
properties of the disordered alloys, and (iii) a quantitative
benchmark to develop and test ab initio based approximations that can
accurately describe the lattice dynamics of disordered solid solutions
and large scale heterostructures, and are computationally more
amenable than direct first-principles calculations.

In this paper, we present first-principles calculations for the
lattice dynamics of three ordered configurations of
Pb(Zr$_{0.5}$Ti$_{0.5}$)O$_3$: the [1:1] ultrathin superlattices
(001), (110) and (111).  In Section II, we give the details of the
computations.  In Section III, we present the relaxed nonpolar
structures, phonon dispersions along selected directions, and detailed
analysis of the unstable modes. Several interesting features not
present in the pure endpoint compounds, and the implications for the
ferroelectric behavior and related properties, are identified and
discussed in Section IV.

\section{Computational Method}

We performed ab initio calculations using density functional theory
(DFT) with the plane wave pseudopotential method.  The vibrational
properties were computed using density functional perturbation theory
(DFPT) \cite{DFPTREV} generalized to ultrasoft
pseudopotentials.\cite{DFPTUS} All calculations were performed using
the PWSCF and PHONON codes.\cite{PWSCF}

We used Vanderbilt ultrasoft pseudopotentials \cite{ultrasoft},
treating as valence states the 3$s$, 3$p$, 3$d$, and 4$s$ electrons of
Ti, the 4$s$, 4$p$, 4$d$, and 5$s$ electrons of Zr, the 5$d$, 6$s$,
and 6$p$ electrons of Pb, and the 2$s$ and 2$p$ electrons of O.  The
inclusion of the semi-core electrons in the valence states is
necessary for an accurate description of these oxides. The exchange
and correlation energy is given within the local density approximation
(LDA) using the parameterization of Perdew and Zunger\cite{CAPZ}.  A
kinetic energy cut-off of 35 Ry was used and the augmentation charges
were expanded up to 350 Ry. For PZT(001), (110) and (111), the
Brillouin zone (BZ) integration was performed using 12 (663 tetragonal
mesh), 9 (446 tetragonal mesh) and 6 (333 fcc mesh) {\bf k}-points in
the irreducible BZ, respectively.

For the prototypical cubic perovskite structure of the pure materials,
PbZrO$_3$ (PZ) and PbTiO$_3$ (PT), we obtained the following values
for the theoretical equilibrium lattice parameter: a(PZ)=7.77 a.u. and
a(PT)=7.37 a.u. (experimental values 7.81 a.u. and 7.50 a.u.) and bulk
modulus B(PZ)=170 GPa and B(PT)=202 GPa.  These compare very well with
results from previous LDA calculations, such as
Ref. \onlinecite{KingSmith}.

To compute the LO-TO splitting at $\Gamma$ arising from the long-range
Coulomb interactions present in the perovskite oxides, it is necessary
to know the electronic dielectric tensor $\epsilon_\infty$ and the
Born effective charge tensors {\bf Z}$^*_s$, where s runs over the
atoms in the unit cell.  These can be computed within the framework of
DFPT, although the current version of PHONON allows these computations
only with norm-conserving pseudopotentials.  For the calculation of
$\epsilon_\infty$ and {\bf Z}$^*_s$ we have therefore used Troullier
and Martins pseudopotentials \cite{Troullier}
with the same valence states as for the 
ultrasoft pseudopotentials. To achieve convergence with the
norm-conserving pseudopotentials an energy cutoff of 80 Ry was needed.
All the other parameters, such as lattice parameters, atomic
positions, and k-point sampling were chosen to be the same as those
optimized with the ultrasoft pseudopotentials.

\section{Results}

The primary focus of our calculations is the eigenfrequencies and
eigenvectors of the unstable polar modes at $\Gamma$. The unstable
eigenvectors of the analytic part of the dynamical matrix are of
particular interest, since they generate symmetry-breaking
energy-lowering distortions of the relaxed structures under conditions
of zero macroscopic electric field.  Some direct connections between
features of the superlattice structure and the unstable modes
emerge. In general, we will see that if pure Ti-O chains are present
in the ordered structure, the lowest frequency unstable modes are
confined in the PT layer and are characterized by displacements
parallel to these chains.  Another connection is between modes of the
superlattice structures and those of strained bulk PT and PZ. This is
especially easy to see in the ideal structure, where the relevant
strain is an isotropic expansion for PT and compression for PZ.  For
the noncubic (001) and (110) structures, the LO-TO splitting gives
rise to a nontrivial anisotropy with changing direction of the
vanishing q vector.  Surprisingly, unstable LO modes are also found to
occur.  Throughout this section, symmetry analysis will be used to
organize and present the results.

\subsection{Relaxed structures}

To find an appropriate high-symmetry reference structure for PZT, we
first placed the atoms as in the ideal cubic perovskite structure with
lattice constant $a$(PZT)=7.57 a.u., obtained by averaging the
theoretical values in the pure endpoint compounds PZ and PT.  This
value is very close to the LDA value of 7.55 a.u. calculated for
PZT(111) by Fornari and Singh.\cite{Fornari} The Zr and Ti atoms are
ordered on the B sites to establish the (001), (110) and (111) [1:1]
superlattices. These superlattice structures will be referred to below
as ``ideal" PZT(001), PZT(110) and PZT(111).  The ``relaxed"
superlattice structures are then obtained by allowing symmetry
preserving relaxations of the internal atomic coordinates, holding the
lattice parameters fixed.  The three ordered supercells each contain
ten atoms, with the Wyckoff labels and corresponding atomic positions
indicating the free internal structural parameters given in
Table~\ref{table1}.

The unit cell of PZT(001) is tetragonal, obtained by doubling the
cubic perovskite cell along the [001] direction (space group
D$^1_{4h}$).  The lattice parameters are fixed at $a$=$a$(PZT) and
$c$=2$a$.  Alternating TiO$_2$ and ZrO$_2$ planes are separated by PbO
planes along the [001] direction, the oxygens in the Wyckoff positions
2f, 2e, and 2g being coplanar with the Ti, Zr and Pb atoms,
respectively.  Pure Ti-O and Zr-O chains run along the [100] and [010]
directions, perpendicular to the direction of compositional
modulation, [001].  The symmetry preserving relaxation of the internal
atomic coordinates consists of displacements along $z$ of the atoms in
the PbO planes (Pb and O(2g) atoms).  In the ideal perovskite
structure, the $z$ values for Pb and O(2g) are exactly $z_{\rm
Pb}=z_{\rm O}=0.25$. After relaxation we find that the PbO planes
shift almost rigidly towards the TiO$_2$ plane with $\Delta z_{\rm
Pb}=-0.20$ a.u.  and $\Delta z_{\rm O}=-0.19$ a.u., resulting in a
lowering of the energy by 291 meV per unit cell.  The relaxed Ti-O(2g)
(Zr-O(2g)) distance is smaller (larger) than the corresponding
distance in the pure compound PT (PZ).  Since the BO$_6$ octahedron
can be regarded as the building block of the perovskite structure, it
is useful to compare the distortion of the TiO$_6$ and ZrO$_6$
octahedra in the supercell relative to the pure endpoint compounds. In
the relaxed (001) structure the TiO$_6$ (ZrO$_6$) octahedron is
stretched (compressed) in the (001) plane, due to the larger (smaller)
lattice parameter ($a$(PT)$<a$(PZT)$<a$(PZ)), and shrinks (expands)
along the [001] direction of compositional modulation. This relaxation
relieves some of the superlattice epitaxial strain by restoring the
local volumes towards their equilibrium values in the bulk endpoint
compounds, though the local c/a ratio is significantly changed.

The PZT(110) structure is also tetragonal (space group D$^1_{4h}$)
with lattice parameters $a$=$\sqrt2$$a$(PZT) and $c$=$a$(PZT).  Along
the [001] direction, TiZrO$_4$ planes alternate with Pb$_2$O$_2$.
Within the TiZrO$_4$ planes the Ti and Zr atoms are placed in a 2D
checkerboard arrangement, each surrounded by four coplanar oxygen
atoms in the 4j position.  Pure Ti-O and Zr-O chains run along the
[001] direction, perpendicular to the two equivalent directions of
compositional modulation, [110] and [1$\overline{1}$0].  The symmetry
preserving relaxation is specified by a single parameter $\Delta x$,
the displacement of the four O(4j) atoms along the Ti-Zr lines in the
TiZrO$_4$ planes.  In the ideal perovskite structure the O atoms are
exactly halfway between Ti and Zr, with $x$=0.25.  After relaxation we
find $\Delta x=-0.125$ a.u., so that the four co-planar O(4j) atoms
surrounding the Ti (Zr) are 0.025 a.u. closer (further) than in PT
(PZ), resulting in a lowering of the energy by 338 meV per unit cell.
Compared to the pure endpoint compounds the TiO$_6$ (ZrO$_6$)
octahedron is stretched (compressed) along the [001] direction, and
shrinks (expands) in the (001) plane of compositional modulation.  As
in PZT(001), this relaxation relieves some of the superlattice
epitaxial strain by restoring the local volumes towards their
equilibrium values in the bulk endpoint compounds, though the local
c/a ratio is significantly changed.

The PZT(111) structure is obtained by placing Ti and Zr atoms on the
simple cubic lattice of B sites in the 3D checkerboard arrangement
corresponding to the rocksalt structure (space group O$^1_{h}$).  The
unit cell is face centered cubic with lattice parameter $a$=2$a$(PZT).
There are no pure Ti-O or Zr-O chains in this structure; in all B-O
chains the Ti and Zr alternate.  The symmetry-preserving relaxation is
specified by a single parameter, $\Delta x$, corresponding to a
breathing of the oxygen octahedra around the Ti and Zr atoms. A
negative value of $\Delta x$ corresponds to a breathing-in
(breathing-out) motion around the Ti (Zr) atoms.  In the ideal
perovskite structure, the O atoms are exactly halfway between Ti and
Zr, with $x=0.25$. After relaxation we find that the oxygen atoms have
moved closer to the Ti, with $\Delta x=-0.115$ a.u., resulting in a
lowering of the energy by 465 meV per unit cell.  The superlattice
TiO$_6$ and ZrO$_6$ oxygen octahedra are remarkably similar to those
in the pure endpoint compounds.  Thus, at least with respect to the
BO$_6$ octahedra (the Pb atoms are fixed), the superlattice strain is
completely relaxed.

\subsection{Z* and $\epsilon_\infty$}

The computed electronic dielectric tensor $\epsilon_\infty$ and the
Born effective charge tensors {\bf Z}$^*_s$ for the relaxed PZT(001)
and PZT(110) structures are given in
Tables~\ref{table2}~and~\ref{table3}.  All matrices are diagonal
except for {\bf Z}$^*_{O(4j)}$ in PZT (110) (note that in PZT(110) the
xy axes are rotated by 45 degrees around z with respect to the
original cubic axes), and therefore, except in that case, only the
diagonal elements are given.

The anisotropy associated with the atomic ordering is generally quite
small. Also, while {\bf Z}$^*_{Zr}$ and {\bf Z}$^*_{Ti}$ are quite
different in the endpoint compounds, even when computed at the same
lattice constant $a$(PZT), the effective charges of Zr and Ti in the
superlattices are very similar.\cite{Bellaiche2} As a result, the
dielectric tensor and effective charges obtained by averaging the
tensors for pure PT and pure PZ, also given in Tables ~\ref{table2}
and~\ref{table3}, are very close to the corresponding quantities for
the superlattices.  In fact, the change in LO phonon frequencies
obtained by using the average values of $\epsilon_\infty$ and {\bf
Z}$^*_s$ instead of the superlattice values is in each case less than
2\%, comparable to the overall accuracy expected for the
first-principles calculation.  The average values might therefore be
expected to be a good approximation in other x=0.5 structures as well.
The results that follow were computed using the average values.

\subsection{Phonon dispersion}

For the three relaxed (001), (110), and (111) PZT superlattices, the
phonon dispersion along the direction of compositional modulation and
the anisotropy at q = 0 are shown in Figure~1.  In this section, we
give an overview of the results, with detailed discussion and analysis
to follow in Sections IIID, E and F.
 
The behavior of the phonon dispersion along the direction of
compositional modulation indicates the degree to which the phonon
vibrations are confined to layers of a particular composition.  This
dispersion is quite flat for most of the modes in the (110) and, to an
even greater degree, in the (001) superlattices. This means that the
corresponding normal modes are strongly localized within atomic planes
perpendicular to the direction of compositional modulation. In
contrast, most of the modes in the (111) superlattice display
significant dispersion, indicating that the normal mode amplitudes are
spread through the supercell. This will be confirmed by the analysis
of individual eigenmodes below.

Due to the long-range character of the Coulomb force, the dynamical
matrix displays non-analytic behaviour in the limit {\bf
q}$\rightarrow$0, resulting in a dependence of the frequencies and
eigenvectors on the direction of the vanishing wave vector, {\bf \^q}.
More precisely, in the long wavelength limit the dynamical matrix can
be written as the sum of an analytic contribution, D$^T$(q=0), and a
non-analytic contribution that depends on the Born effective charge
tensors {\bf Z}$^*_s$ and the electronic dielectric tensor
$\epsilon_\infty$:
\begin{eqnarray}\nonumber
{\bf D}_{s s^\prime}({\bf q}\rightarrow 0) 
 &=& {\bf D}_{s s^\prime}^{\rm T} (q=0)                        \\ 
 &+& \frac {4\pi e^2} {\Omega\sqrt{M_sM_{s^\prime}}} \
 \frac { ({\bf Z}^*_s \cdot {\bf \hat q}) 
         ({\bf \hat q} \cdot {\bf Z}^*_{s^\prime}) }
   { {\bf \hat q} \cdot 
     \epsilon_\infty \cdot {\bf \hat q} },
\end{eqnarray}
where D$^T$(q=0) is the dynamical matrix computed with the boundary
condition of zero macroscopic electric field ({\bf E}=0), M$_s$ is the
atomic mass of the atom $s$, and $\Omega$ is the volume of the unit
cell.\cite{BornHuang,CochranCowley} In the lefthand panels of the
three plots in Figure~1, we show how the frequencies
depend on the angle $\theta$ between the direction of the vanishing
wave vector, $\hat {\bf q}$, and the direction of compositional
modulation.  For a cubic structure, such as the (111) superlattice,
the limiting frequencies do not depend upon the angle $\theta$ and can
simply be identified as LO (longitudinal optic) and TO (transverse
optic), though as in the perovskite structure, with multiple optic
modes of the same symmetry, no one-to-one correspondence can be made
between individual TO and LO modes\cite{Zhong}.  For the tetragonal
(001) and (110) superlattices, the limiting frequencies do depend on
the angle $\theta$.  This gives rise to bands of LO frequencies which
can be identified by their dispersion in Figure~1 and
which will be discussed in detail in Section IIIF.

All three superlattices display lattice instabilities; in fact, their
phonon spectra contain entire branches of imaginary frequencies.  The
imaginary frequencies are indicated by the negative values in Figure~1
and the corresponding region of unstable modes is shaded in gray.  The
most unstable mode is observed at $\Gamma$ in all three cases,
consistent with our expectation that as for bulk PZT, the ground
state of these superlattices should be ferroelectric.  However, (111)
is anomalous in that this dominant $\Gamma$ mode is actually a
nonpolar pure oxygen mode that will not produce a ferroelectric ground
state, as will be discussed in more detail below.  In the following,
we focus on the behavior at the zone center, with particular attention
to the unstable modes.

\subsection{Unstable eigenmodes of D$^T$(q=0)}

Diagonalization of the analytic contribution to the dynamical matrix,
D$^T$(q=0), gives the energies of the normal modes in zero macroscopic
field. These modes can be classified using the full symmetry of the
point group of the space group, as will be described in detail below.
The transverse IR-active phonons (TO) and the non-polar phonons are
eigenmodes of D$^T$(q=0) as well as of the full dynamical matrix,
while the LO modes are mixed by the nonanalytic contribution, as
will be seen in Section F.

Since D$^T$(q=0) has the full symmetry of the point group of the space
group, we can classify its eigenmodes according to the irreducible
representations (irreps) of D$_{4h}$ for PZT(001) and PZT(110) and
O$_{h}$ for PZT(111), following the nomenclature of
Ref. \onlinecite{grouptheory}.  These labels specify the polar
character of the modes, which atoms are involved in the vibration, and
the directions of the corresponding atomic displacements.

For the (001) superlattice, the eigenmodes are labelled by five irreps:
\[
8\Gamma_{5^\prime}^{(2)} \oplus 2\Gamma_5^{(2)} \oplus 
6\Gamma_{1^\prime} \oplus 
2\Gamma_1 \oplus 2\Gamma_{3^\prime}.
\]
where the coefficient gives the multiplicity of the irrep and the
superscript indicates its dimension.  The two-dimensional irreps
$\Gamma_{5^\prime}$ and $\Gamma_5$ correspond to vibration in the
xy-plane perpendicular to the direction of compositional modulation
($\perp$).  The others are one-dimensional irreps and correspond to
vibrations along the direction of compositional modulation
($\parallel$) .  There are twelve IR-active polar TO modes (seven
$\Gamma_{5^\prime}(\perp)$ and five $\Gamma_{1^\prime}(\parallel)$ modes) 
in which all
of the atoms participate.  The remaining $\Gamma_{5^\prime}$ mode and
$\Gamma_{1^\prime}$ mode are the zero-frequency acoustic modes.  All
the other modes are non-polar: the two $\Gamma_{3^\prime}$ modes
involve only O(2e) and O(2f) z-vibrations, the two $\Gamma_1$ modes
involve only O(2g) and Pb z-vibrations, and the two $\Gamma_5$ modes
involve only O(2g) and Pb xy-vibrations.

For the (110) superlattice, the eigenmodes are labelled by nine irreps:
\[
8\Gamma_{5^\prime}^{(2)} \oplus \Gamma_5^{(2)} \oplus 6\Gamma_{1^\prime} 
\oplus \Gamma_{3^\prime} \oplus \Gamma_{4^\prime} \oplus \Gamma_1
\oplus \Gamma_2 \oplus \Gamma_3 \oplus \Gamma_4.
\]
These irreps correspond to vibrations in the xy-plane of compositional 
modulation ($\parallel$) or to vibrations in the perpendicular 
z direction ($\perp$).  
As in the (001) supercell there are twelve IR-active polar TO modes 
(seven $\Gamma_{5^\prime}(\parallel)$ and five $\Gamma_{1^\prime}(\perp)$)
in which all of the atoms participate.
The remaining $\Gamma_{5^\prime}$ mode and $\Gamma_{1^\prime}$ mode
are the zero-frequency acoustic modes.  All the other modes are
non-polar: the $\Gamma_{3^\prime}$ mode involve only Pb z-vibrations,
the $\Gamma_{4^\prime}$ and $\Gamma_5$ modes involve only O(4j)
z-vibrations, and the $\Gamma_1$, $\Gamma_2$, $\Gamma_3$, $\Gamma_4$
modes also involve only O(4j) vibrations but in the xy plane.

For the (111) supercell the eigenmodes are labelled by nine irreps:
\[
5\Gamma_{15}^{(3)} \oplus 2\Gamma_{25^\prime}^{(3)} \oplus 
\Gamma_{25}^{(3)} \oplus \Gamma_{15^\prime}^{(3)} \oplus 
\Gamma_{12}^{(2)} \oplus \Gamma_1.
\]
There are four IR-active polar TO modes, $\Gamma_{15}$, which involve
atomic displacements of all the atoms along x, y or z.
The remaining $\Gamma_{15}$ mode is the acoustic zero frequency mode.
All the other modes are non-polar:
the two $\Gamma_{25^\prime}$ modes are the odd linear combination
of the two Pb(x,y,z) and one oxygen displacement pattern, and
the remaining $\Gamma_{25}$, $\Gamma_{15^\prime}$,
$\Gamma_{12}$, and  $\Gamma_1$ are all unique pure oxygen modes.

For both the ideal and relaxed structures of the three superlattices,
the eigenfrequencies and symmetry labels of the {\it unstable}
eigenmodes of D$^T$(q=0) are given in Table~\ref{table4}.  A
comparison of the instabilities of the three ideal superlattice
structures provides information on their sensitivity to the atomic
configuration. In the ideal structures the atomic positions are
identical and the only difference is in the arrangement of Zr and Ti
atoms on the B sites.  It can immediately be seen that the unstable
eigenfrequencies are quite sensitive to the arrangement of atoms on
the B site. In part, this reflects the different displacement patterns
imposed by symmetry. Even if the Zr and Ti atoms were completely
equivalent, these lists would differ because of the different q-points
of the primitive Brillouin zone that are folded into the smaller cubic
Brillouin zone of each supercell ($\Gamma$ and X for (001), $\Gamma$
and R for (111), and $\Gamma$ and M for (110)), so only half of the
modes (those from primitive $\Gamma$) would match for each pair. With
the actual atoms in place the mixing of the different ``folded'' mode
combinations gives quite different results in the three superlattices.

For each superlattice structure, the effect of atomic relaxation has
been investigated by comparing frequencies of the ideal structure with
those of the relaxed structure. In PZT(001) the shift in frequencies
is substantial, and in some cases even leads to a reordering of the
modes (for example, the unstable $\Gamma_{3^\prime}$ and
$\Gamma_{5^\prime}$ modes). The relaxations in the other two
superlattice structures cause smaller changes.  The larger shift in
PZT(001) may be due to the larger atomic relaxations observed in this
structure, which involve Pb displacements in addition to the oxygen
displacements.

The unstable polar TO modes will be described in detail in the next
section.  In addition, each of the three superlattice structures has
one pure oxygen instability.  The oxygen instability in PZT(001), the
$\Gamma_{3^\prime}$ mode, is mainly confined in the PZ layers and
corresponds to the $\Gamma_{25}$ mode of bulk PZ. In the superlattice
the PZ layers are under compressive strain ($a$(PZT)$<a$(PZ)) and the
origin of this oxygen instability is in the pressure dependence of the
$\Gamma_{25}$ mode of PZ which becomes unstable under compression.  We
note that this $\Gamma_{3^\prime}$ oxygen instability is very
sensitive to strain. In fact, the atomic relaxation alone causes a
large shift of its eigenfrequency.  The oxygen instability in
PZT(110), the $\Gamma_{2}$ mode, is a rotation of the oxygen octahedra
around the [001] direction and corresponds to an M mode of the
primitive bulk cubic cell.  In PZT(111) the oxygen instability is the
most unstable $\Gamma_{25}$ mode. It involves the rotation of the
oxygen octahedra around the $<111>$ directions and corresponds to the
bulk R$_{25}$ mode. This unstable mode contributes to the
orthorhombic antiferroelectric ground state of pure PZ and is
responsible for the tetragonal antiferrodistortive ground state of
SrTiO$_3$.  
Comparing the instabilities of PZT(111) with those of PZT(001) and
PZT(110), we see that the frequency of the oxygen instability is 
similar in all three systems. However, in PZT(111) the most unstable
polar mode has been significantly shifted upwards, which we attribute
to the lack of intact Ti-O chains in any direction. This leaves the
oxygen instability as the dominant unstable mode.

\subsection{Confined TO polar instabilities}

Each of the three superlattice structures possesses TO polar
instabilities, indicated in boldfaced type in Table~\ref{table4}, that
can generate candidate ferroelectric structures.  PZT(001) and
PZT(110) exhibit ``two-mode behavior," that is, there are two types of such
unstable modes, one with atomic displacements mainly confined in the
PT layer and the other in the PZ layer.  The atomic displacements of
the two most unstable TO modes of relaxed PZT (001), $\Gamma_{5^\prime}$ and
$\Gamma_{1^\prime}$, are shown in Figure~2. The $\Gamma_{5^\prime}$ is
a PT-like mode polarized parallel to the interfaces; it is confined in
the PT layer and is remarkably similar to the unstable TO mode of bulk
PT (also shown in Figure~2 for comparison).  The $\Gamma_{1^\prime}$
mode, on the other hand, is a PZ-like mode polarized perpendicular to
the interfaces; it is mainly confined in the PZ layer and is
remarkably similar to the unstable TO mode in bulk PZ. The confinement
of these modes in the PT and PZ layers is consistent with their flat
dispersion for {\bf q} perpendicular to the interfaces.

The atomic arrangements and projections in the (110) and (111)
structures do not lend themselves well to representation in a figure
analogous to Figure~2.  A convenient way of characterizing the
confinement of the polar instabilities in all three superlattice
structures is to decompose the mode effective charge, $\overline{\bf
Z}$, into two contributions: one from the Zr-centered cell and one
from the Ti-centered cell: $\overline{\bf Z}=\overline{\bf Z}_{\rm
Zr}+\overline{\bf Z}_{\rm Ti}$.  These local contributions to the mode
effective charge, which will be referred to as ``local effective
charges," are defined as
\begin{equation}
\overline{\bf Z}_{\rm Ti(Zr)} = \sum_{s,\beta} w_{s} Z^*_{\alpha \beta ;s}
                          u_{\beta ,s} 
\end{equation}
where the sum over the atoms, $s$, runs only over the atoms in the Ti
(Zr) centered cell and the weighting factor $w_{s}$ is $w$=1 for Ti
(Zr), $w$=0.5 for the 6 oxygen atoms and $w$=0.125 for the 8 Pb atoms.
In the case that the modes are confined, we expect the local effective
charge to have its largest value in the corresponding layer.

The local effective charges for PZT(001) have been included in
Figure~2. For the PT-like mode, $\Gamma_{5^\prime}$,
$\overline{Z}_{\rm Ti}$=7.29 is more than three times larger than
$\overline{Z}_{\rm Zr}$, indicating a high degree of confinement.
Moreover, it is very close to the mode effective charge of the
unstable TO mode in PT (at the equilibrium lattice constant),
$\overline{Z}$=7.57, confirming the local similarity to the bulk PT
unstable mode.  For the PZ-like mode, $\Gamma_{1^\prime}$, the larger
contribution is from the Zr-centered cell, $\overline{Z}_{\rm
Zr}$=4.93, which is likewise very close to the mode effective charge
of the unstable TO mode in PZ, $\overline{Z}$=4.86.

The local effective charges of the unstable TO modes of all three
superlattice structures are given in Table~\ref{table5}. By symmetry,
the induced polarizations are parallel to the direction of atomic
displacement for each mode, also included in the table.  For almost
every unstable TO mode of the (001) and (110) superlattices, either
$\overline{Z}_{\rm Ti}$ or $\overline{Z}_{\rm Zr}$ strongly dominates,
indicating that the polar instabilities are mainly confined either to
the PT or to the PZ layers, respectively.  The confined character of
the modes is seen to be rather insensitive to the relaxation.  Just as
for the (001) case discussed above, the most unstable mode for
PZT(110) is a PT-like mode with polarization aligned in the PT plane
(perpendicular to the direction of compositional modulation, $\perp$)
followed by a PZ-like mode with polarization parallel to the direction
of compositional modulation, $\parallel$.

In contrast to the two-mode behavior of PZT(001) and PZT(110),
PZT(111) exhibits one-mode behavior. For its single unstable TO mode,
the local effective charges are almost identical and the polar
instability is distributed throughout the supercell. The different
behavior of PZT (111) is a consequence of the fcc symmetry; in fact,
the displacements of the six oxygens in the supercell are
symmetry-related, so that any mode in which oxygen displacements are
allowed will in general involve the whole supercell, preventing 
the confinement to occur.

There are two distinct factors that could produce the observed mode
confinement: the difference between the atomic mass of the two species
occupying the B sites (Ti and Zr) or the dependence of the interatomic
force constants upon atomic configuration (i.e. the atomic force
constants may depend upon the chemical identity of either of the two
interacting atoms and upon the surrounding atomic configuration). In
some semiconductor heterostructures, such as
AlAs/GaAs\cite{AlGaAs1,AlGaAs2}, the interatomic force constants
(IFCs) are to a good approximation unaffected by the substitution,
Al/As, and the observed character of the heterostructure modes is the
result of the difference in mass.  One might hope that the
isoelectronic substitution Ti/Zr could be similarly understood.  To
check this, the eigenfrequencies and local effective charges of the
dynamical matrices in which the masses of Ti and Zr have been replaced
by their average are shown in Table~\ref{table6}.  Changing the value
of the masses results in a frequency shift that is larger for those
modes that involve larger displacements of the Ti or Zr ions. In
particular the PT-like modes of (001) and (110) are shifted to higher
frequency.  The PZ-like modes are almost unaffected since the atomic
displacements of Zr and Ti are almost zero in these modes. By the
analysis of the local effective charges in Table~\ref{table6} it is
evident that the mode confinement is still present when Ti and Zr have
the same mass. It must therefore be that the IFCs depend strongly upon
atomic configuration, with interactions involving Zr atoms
significantly differing from their Ti counterparts. The important
consequence of this observation is that approximations for the IFCs
that neglect the effect of compositional modulation, with the most
widely used being the virtual crystal approximation, cannot adequately
capture the character of the instabilities in PZT.

\subsection{LO-TO splitting and anisotropy at q=0}

The effect of adding the non-analytic term to the dynamical matrix is
to mix the IR-active modes that have a non-vanishing longitudinal
component of the polarization thereby generating longitudinal
IR-active modes (LO).  This results in a splitting of the LO and TO
frequencies for vanishing q vector that, in general, depends upon the
direction (as noted in the angular dispersions of figure 2).

\begin{figure}
  \vbox{ 
        \centerline{ 
         \psfig{figure=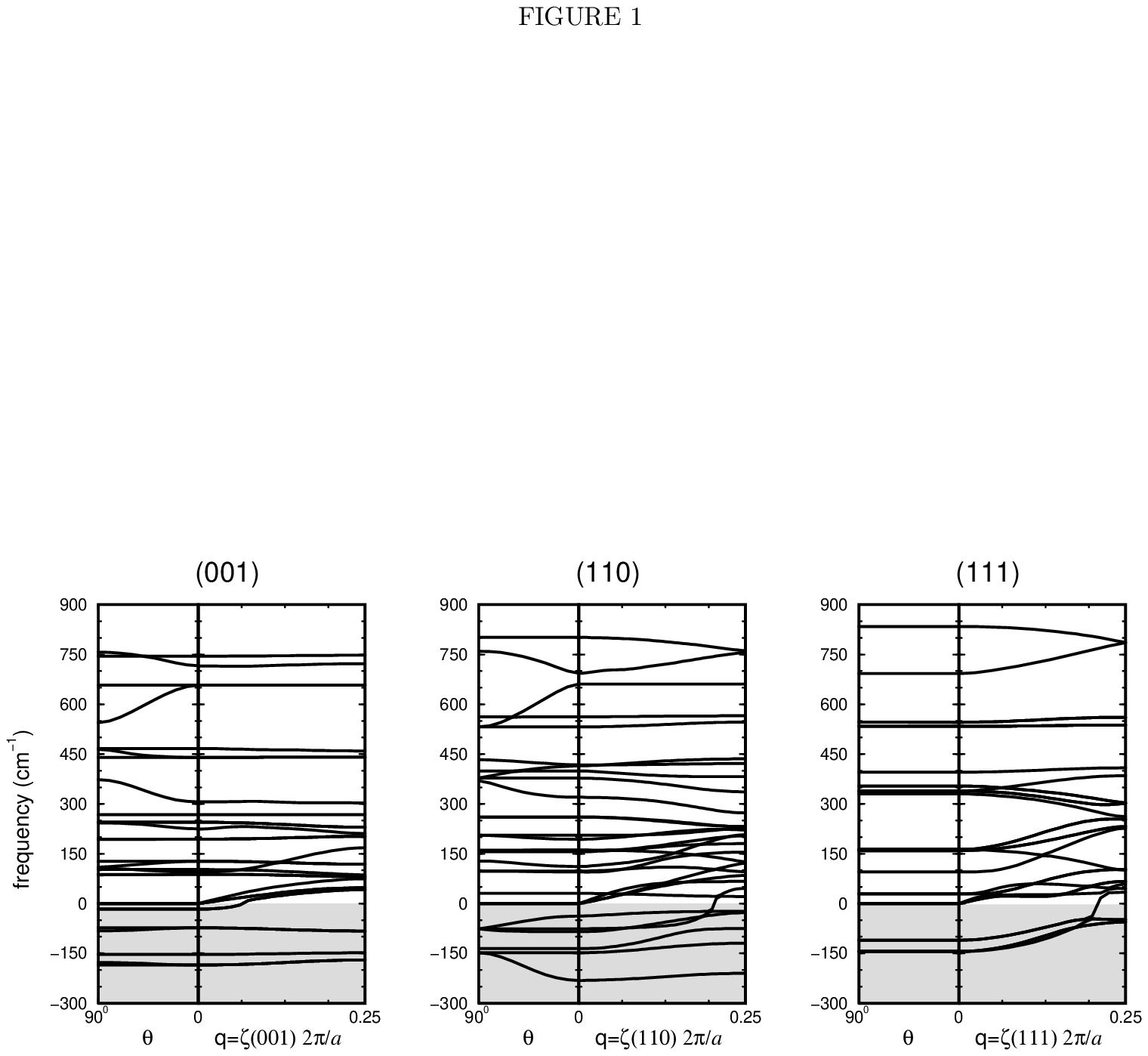,width=16cm} 
         }
  \caption{Phonon dispersion along the direction of compositional modulation and q=0 anisotropy 
   for the three relaxed PZT superlattice structures.
}
}
\label{dispersion}
\end{figure}

The angular dispersions of the LO modes have been already displayed in
Figure~\ref{dispersion}.  For a more detailed discussion, and to
assist in the interpretation of the figure, the frequencies of all the
TO and LO modes of the three relaxed supercells are given in
Table~\ref{table7} with their symmetry labels and mode effective
charges.  For the (001) and (110) superlattices the LO frequencies are
direction dependent and we give their values for $\hat{\bf q}$
parallel and perpendicular to the direction of compositional
modulation in each supercell, corresponding to $\theta=0$ and
$\theta=90\deg$ in the angular dispersion shown in Figure~1,
respectively.  For the cubic (111) superlattice, symmetry
considerations show that the LO frequencies do not depend upon
$\hat{\bf q}$.

Examination of the table shows that the (001) and (110) superlattices
have unstable LO modes, a feature not present in the pure compounds.
These modes are not confined, showing significant relative motion of
the cations and anions in both the Zr and Ti centered subcells of the
supercell.  However, the induced polarizations in the two subcells are
opposite in direction, and resulting near cancellation makes the
unstable LO mode effective charges much smaller than those associated
with the TO instabilities.

\section{Discussion}

As for the pure endpoint compounds, the unstable TO phonons are
expected to be valuable indications of the existence and nature of a
ferroelectric ground state for the PZT superlattices here considered,
as they generate energy-lowering distortions with nonzero spontaneous
polarization P$_s$ under conditions of zero macroscopic electric field
(${\bf E}=0$).

With dominant TO instabilities, PZT(001) and PZT(110) are virtually
guaranteed to have ferroelectric ground states. In previous
first-principles studies, attention has focused on ferroelectric
structures of PZT(001) with polarization along the [001]
direction. The two-mode behavior discussed here suggests that the true
ground state could be both more complicated and more interesting.
Freezing in the unstable TO phonons of the (001) superlattice give
rise to a macroscopic polarization that is parallel to the interfaces
in the PT layers for $\Gamma_{5^\prime}$ and perpendicular to the
interfaces in the PZ layers for $\Gamma_{1^\prime}$. Depending on the
anharmonic coupling between these modes and their coupling to
homogeneous strain, either or both of these modes could freeze in to
give a ground state structure with polarization along the [001]
direction (the c phase), along [100] (the a phase), along [110] (the
aa phase), in the (010) plane (the ac phase), or in the (110) plane
(the designations of the phases correspond to those used in
Ref. \onlinecite{Pertsev} in a different context).

The choice of equilbrium structure type from the list above should
also be quite sensitive to the mechanical and electrical boundary
conditions.  Applied stresses, epitaxial constraints and applied
electric fields will result in relative shifts of the two modes and
changes in their coupling, and could result in transitions from one
phase to another.  In particular, this system might provide a
microscopic realization of a large-strain electrostrictive actuator as
recently suggested by Bhattacharya. \cite{Bhattacharya}

In contrast, PZT(111) has a dominant nonpolar instability, with the
ferroelectric TO mode slightly higher in frequency.  The ground state
has been analyzed in detail by Fornari and Singh\cite{Fornari}.  They
observe, consistent with our results, that the ferroelectric and
rotational distortions are quite competitive in energy and that this
competition is very sensitive to strain.

The (001) and (110) supercells display LO as well as TO instabilities,
a feature not present in the pure compounds.  This presents the
interesting possibility of a perpendicularly polarized ferroelectric
ground state structure for a freestanding PZT film or nanoparticle.
The energy of such a structure, with a depolarizing field ${\bf
E}=-4\pi{\bf P}$, would in general be prohibitively high,
corresponding to the stiffness of the LO mode.  However, with an
unstable LO mode, there would be a net energy gain despite the
depolarizing field. Since Z* is rather low for these modes, the
spontaneous polarization would presumably be quite small, though still
possibly observable or even useful. Whether this instability can be
made to yield an actual ground state through application of
appropriate mechanical boundary conditions remains to be investigated.

Aside from the physical implications of the calculated phonons, these
results can also be used as a test of approximate schemes for
obtaining the phonons of these and larger ordered superstructures.  It
is clear that while the virtual crystal approximation has been seen to
work well for PZT(111), the requisite one-mode behavior of this
structure is actually quite atypical.  Generalizing from the lower
symmetry PZT(001) and (110) structures, we expect that larger
superstructures, with even lower symmetry, will exhibit multi-mode
behavior.  A key ingredient in reproducing this behavior is that
approximations should allow for force constants involving Ti
displacements to be different from the analogous force constants
involving Zr displacements.  The development and application of a
suitable method, which was proposed in Ref. \onlinecite{Comparative},
will be the subject of a future publication.\cite{mixing}

\section{Summary}

We have presented an ab initio study of the lattice dynamics of the
high-symmetry nonpolar reference structures of three ordered
configurations of PZT: the three ultrathin PT/PZ [1:1] superlattices
(001), (110) and (111).  In particular we focused on the zone-center
lattice instabilities that can generate symmetry-breaking
energy-lowering distortions and eventually lead to a phase transition.
In PZT(001) and (110), these show a two-mode behavior that may have
interesting consequences for ferroelectricity in these systems.  Some
direct connections between features of the superlattice structure and
the unstable modes emerge. In general, if pure Ti-O chains are present
in the ordered structure, the lowest frequency unstable modes are
confined in the PT layer and are characterized by displacements
parallel to these chains.  Another connection is between modes of the
superlattice structures and those of strained bulk PT and PZ.
Finally, we have verified that the character of the modes is not
simply due to the small difference between the atomic mass of the
intermixing cations, Ti and Zr, as is the case in some semiconductors,
such as (Al,Ga)As, but that the effects of compositional modulation on
the IFCs play an important role.  Consequently, simple approximations
for the IFC that neglect the effects of alloying, such as the virtual
crystal approximation cannot provide an accurate description of the
lattice dynamics in these systems.

\acknowledgments

This work was supported by ONR N00014-00-1-0261. The work of
K.M.R. was performed in part at the Aspen Center for Physics. The
majority of the computations were performed at the Maui High
Performance Computing Center.

\newpage

\onecolumn

\twocolumn

\begin{figure}
  \vbox{ 
    \centerline{
      \psfig{figure=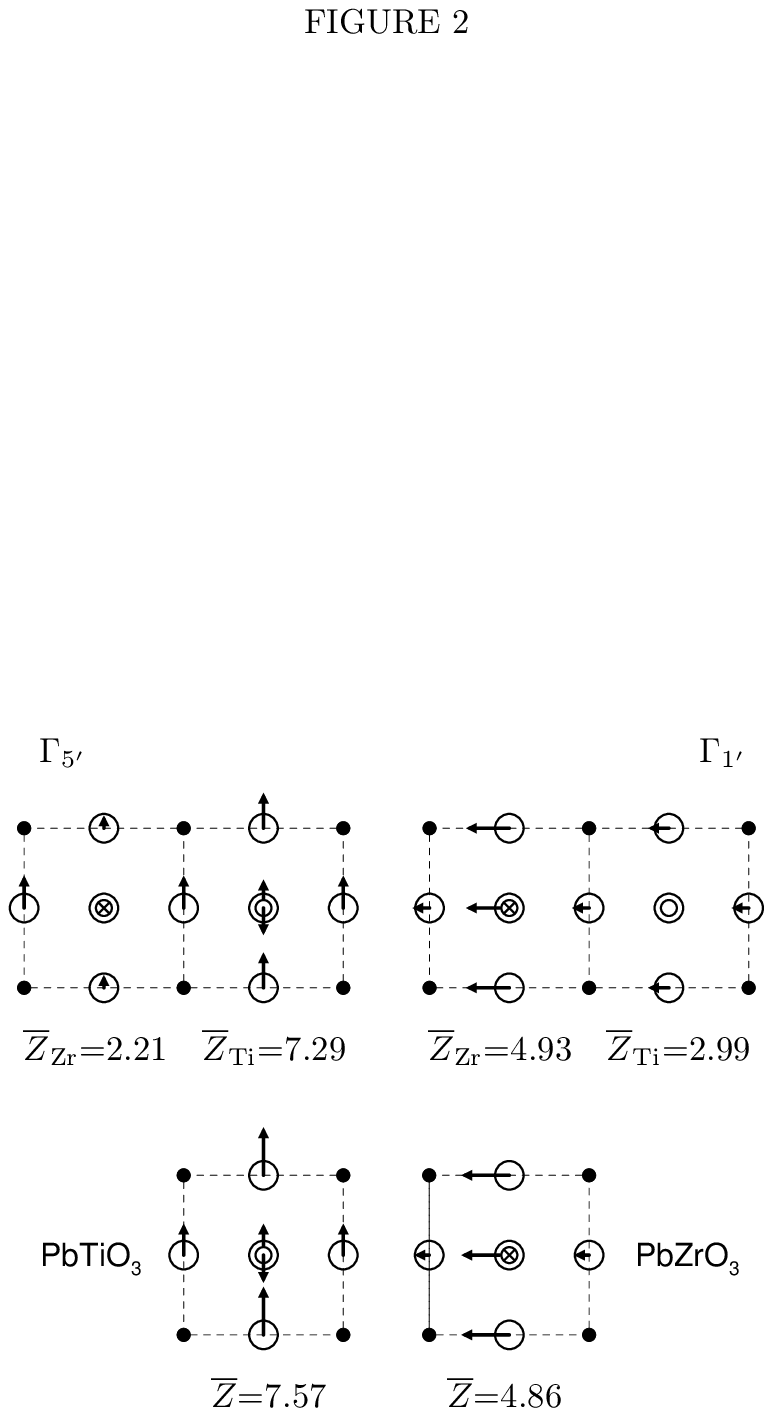,width=8cm}
      }
  \caption{Atomic displacements for the two most unstable TO polar
modes in PZT(001), $\Gamma_5^\prime$ and $\Gamma_1^\prime$, and the
TO1 unstable modes of bulk PbTiO$_3$ and PbZrO$_3$. The projection in
the (100) plane is shown. The different atomic species are represented
as follows: small black circles for Pb, large white circles for O,
small white circles for Ti, and small white circles with a cross for
Zr. Note that the Zr displacement is close to zero in all modes shown and the
atomic vibration in the Zr-centered cell is dominated by the
displacements of the oxygen atoms.
}
}
  \label{relaxed001modes}
\end{figure}

\newpage

\begin{table}
\caption{
Wyckoff labels and atomic positions, in reduced coordinates, 
for the three superlattice structures.
For the (111) superlattice structure, the reduced coordinates are
in terms of the lattice vectors of the conventional fcc unit cell.
For Wyckoff positions with multiplicity greater than one, the first
atom in the list will be referred to as the ``representative atom."
}
\begin{tabular}{c c c }
 PZT (001) &
   Pb(2h)   & (0.5 0.5  $\pm z_{\rm Pb}$) \\
 & Ti(1a)   & (0   0  0) \\
 & Zr(1b)   & (0   0  0.5) \\
 & O(2g)    & (0    0 $\pm z_{\rm O}$) \\
 & O(2f)    & (0    0.5  0), (0.5 0 0) \\ 
 & O(2e)    & (0    0.5  0.5), (0.5 0 0.5) \\
\hline
 PZT (110) &
     Pb(2e) & (0.5 0    0.5),  (0.5 0 0.5)    \\
 &   Ti(1a) & (0   0  0) \\
 &   Zr(1c) & (0.5 0.5 0) \\
 &   O(4j)  & ($\pm x$ $\pm x$ 0) \\
 &   O(1b)  & (0 0 0.5) \\
 &   O(1d)  & (0.5 0.5 0.5) \\
\hline
 PZT (111) &
     Pb(2c)   & $\pm$(0.25 0.25 0.25) \\ 
 &   Ti(1a)   &  (0   0  0) \\
 &   Zr(1b)   & (0.5 0.5 0.5) \\
 &    O(6e)   & ($\pm x$ 0 0), (0 $\pm x$ 0), (0 0 $\pm x$)\\ 
    \end{tabular}
    \label{table1}
\end{table}


\begin{table}
\caption{
Electronic dielectric tensor and Born effective charge tensors for 
representative atoms of the relaxed PZT(001) structure.
Only the diagonal elements are given; the off-diagonal elements are zero 
by symmetry. The dielectric and Born effective charge tensors
obtained by averaging the tensors computed for pure PT and pure PZ (at 
the average lattice parameter a(PZT)) are given on the right hand side 
for comparison.}
\begin{tabular}{c r r r | r r r}
      &\multicolumn{3}{c}{PZT(001)}&\multicolumn{3}{c}{average}\\
\hline
   Pb  
   & 3.91  &  3.91  &  3.87     &      3.90 & 3.90 & 3.90  \\
   Ti
   & 6.51  &  6.51  &  6.36     &      6.47 & 6.47 & 6.47   \\
   Zr
   & 6.33  &  6.33  &  6.40     &      6.47 & 6.47 & 6.47   \\
   O(2g)
   &-2.54  & -2.54  & -5.20     &     -2.54 &-2.54 &-5.29   \\
   O(2f)
   &-2.77  & -5.33  & -2.28     &     -2.54 &-5.29 &-2.54   \\
   O(2e)
   &-2.36  & -5.12  & -2.78     &     -2.54 &-5.29 &-2.54   \\
\hline
$\epsilon_\infty$ & 7.31 & 7.31 & 7.30  &  7.61 & 7.61 & 7.61\\
    \end{tabular}
    \label{table2}
\end{table}


\begin{table}
\caption{Electronic dielectric tensor and Born effective charge tensors 
for the relaxed PZT(110) structure. 
Only the diagonal elements are 
given, except for the O(4j) atoms; the off-diagonal elements are zero by 
symmetry. The dielectric tensor and Born effective charges
obtained by averaging the tensors computed for pure PT and pure PZ (at
the average lattice parameter a(PZT)) are given on the right hand side 
for comparison. Note that in the (110) supercell the axis are rotated by 
45 degrees around $z$ with respect to the original cubic axes.}
\begin{tabular}{c  c | c }
      & PZT(110) & Average\\
\hline
   Pb   & 
   \begin{tabular}{ccc} 4.24  &  3.62  &  3.93 \end{tabular} &      
   \begin{tabular}{ccc} 3.90  &  3.90  &  3.90 \end{tabular}  \\
   Ti   & 
   \begin{tabular}{ccc} 6.47  &  6.47  &  6.80 \end{tabular} &
   \begin{tabular}{ccc} 6.47  &  6.47  &  6.47 \end{tabular}  \\
   Zr   & 
   \begin{tabular}{ccc} 6.47  &  6.47  &  6.24 \end{tabular} &  
   \begin{tabular}{ccc} 6.47  &  6.47  &  6.47 \end{tabular} \\
   O(1b)& 
   \begin{tabular}{ccc} -2.49  & -2.49  & -5.48 \end{tabular} &  
   \begin{tabular}{ccc} -2.54  &-2.54   & -5.29 \end{tabular} \\
   O(1d)& 
   \begin{tabular}{ccc} -2.61  & -2.61  & -5.07 \end{tabular} &  
   \begin{tabular}{ccc} -2.54  & -2.54  & -5.29 \end{tabular}  \\
   & & \\
   O(4j)& 
    $\left( \begin{tabular}{ccc} 
       -3.90 & -1.36 & 0 \\ -1.36 & -3.90 & 0 \\ 0 & 0 & -2.58
   \end{tabular} \right)$  &
   $\left( \begin{tabular}{ccc} 
       -3.92 & -1.38 & 0 \\ -1.38 & -3.92 & 0 \\ 0 & 0 & -2.54
   \end{tabular}  \right)$\\
   & & \\
\hline
$\epsilon_\infty$ & 
   \begin{tabular}{ccc} 7.31 & 7.31 & 7.41 \end{tabular} &  
   \begin{tabular}{ccc} 7.61 & 7.61 & 7.61 \end{tabular} \\
    \end{tabular}
    \label{table3}
\end{table}

\begin{table}
\caption{
Frequencies and symmetry labels of the unstable eigenmodes of D$^T$(q=0) 
for the relaxed and ideal 
PZT(001), (110) and (111) structures. The values in boldfaced type correspond 
to IR-active TO modes, the others being nonpolar modes.
In PZT(001) and PZT(110) the corresponding atomic displacements
are either parallel ($\parallel$) or perpendicular ($\perp$) to the direction 
of compositional modulation, as indicated. Frequencies are in cm$^{-1}$.
}
\begin{tabular}{l c c c| l c c c | l c c }
     & \multicolumn{3}{c}{(001)}  &
     & \multicolumn{3}{c}{(110)}  &
     & \multicolumn{2}{c}{(111)}  \\   
&  & \multicolumn{1}{c}{relaxed}  & \multicolumn{1}{c}{ideal}&
&  & \multicolumn{1}{c}{relaxed}   & \multicolumn{1}{c}{ideal} &  
  & \multicolumn{1}{c}{relaxed} &   \multicolumn{1}{c}{ideal}         \\
\hline
$\Gamma_{5^\prime}$ & $ \perp$     &{\bf -185}& {\bf -235}& 
 $\Gamma_{1^\prime}$ & $ \perp$     &{\bf -231} & {\bf -240}& 
  $\Gamma_{25}$ & -144    & -155\\
$\Gamma_{1^\prime}$ & $ \parallel$ &{\bf -177}& {\bf -151}& 
  $\Gamma_{5^\prime}$ & $ \parallel$ &{\bf -148} & {\bf -158}& 
   $\Gamma_{15}$ & {\bf -110} & {\bf -115}\\
$\Gamma_{3^\prime}$ & $ \parallel$ &-153 &  -72 & 
  $\Gamma_{2}$ &$\parallel$ &-134  & -147 &    &         &     \\
$\Gamma_{5^\prime}$ & $\perp$     & {\bf -73}& {\bf -114}& 
  $\Gamma_{5^\prime}$ & $ \parallel$ & {\bf -76} &  {\bf -71}&   &   &   \\
$\Gamma_{5}$ & $\perp$     & -17 &  -22 & 
  $\Gamma_{1^\prime}$ & $ \perp$   & {\bf -37} &  {\bf -67}&    &      &     \\
    \end{tabular}
    \label{table4}
\end{table}


\begin{table}[h]
\caption{The mode effective charge tensors, $\overline{\bf Z}$, 
associated with the 
unstable IR-active TO modes are decomposed into two contributions,
one from the Zr-centered cell and one from the Ti-centered cell:
$\overline{\bf Z}=\overline{\bf Z}_{\rm Zr}+\overline{\bf Z}_{\rm Ti}$. 
Results are shown
for the relaxed and ideal PZT(001), (110) and (111) structures. 
The mode effective charges for the unstable TO modes of bulk PbZrO$_3$
and PbTiO$_3$ are given for comparison, both at the equilibrium 
volume and strained to the average volume of the 50\% alloy, $V=a^3$(PZT).
}
\begin{tabular}{ccccc|cc}
 superlattice & & &\multicolumn{2}{c|}{relaxed} &  \multicolumn{2}{c}{ideal} \\
           & & &$\overline{Z}_{\rm Zr}$ & $\overline{Z}_{\rm Ti}$ & 
              $\overline{Z}_{\rm Zr}$ & $\overline{Z}_{\rm Ti}$ \\
 \hline
 (001)
 &$\Gamma_{5^\prime}$ &($\perp$)    & 2.21  & 7.29   &  1.57  &  8.11   \\
 &$\Gamma_{1^\prime}$ &($\parallel$)& 4.93  & 2.99   &  4.44  &  3.71   \\
 &$\Gamma_{5^\prime}$ &($\perp$)    & 3.38  &-1.56   &  3.76  & -0.65   \\
 &                    &             &       &        &        &         \\
 (110) 
 &$\Gamma_{1^\prime}$ &($\perp$)    & 2.26  & 7.73   &  2.19  &  7.87   \\
 &$\Gamma_{5^\prime}$ &($\parallel$)& 4.38  & 2.81   &  4.11  &  2.91   \\
 &$\Gamma_{5^\prime}$ &($\parallel$)& 0.13  & 3.03   &  0.60  &  3.42   \\
 &$\Gamma_{1^\prime}$ &($\perp$)    & 2.74  & 0.04   &  3.10  & -0.06   \\
 &                    &             &       &        &        &         \\
 (111)
 &$\Gamma_{15}$       &             & 3.81  & 3.81   &  3.71  &  4.00   \\
 &                    &             &       &        &        &         \\
 bulk & & &\multicolumn{2}{c|}{equilibrium} &  \multicolumn{2}{c}{strained} \\
\hline
 PbZrO$_3$   &        &             & 4.86  &        &  4.87  &         \\
 PbTiO$_3$   &        &             &       & 7.57   &        &  8.00   \\
\end{tabular}
    \label{table5}
\end{table}

\begin{table}[h]
\caption{ The frequencies and local effective charges
obtained for the relaxed superlattice structures when 
the Ti and Zr atoms are assumed
to have the same atomic mass: 
M$_{\rm Ti}$=M$_{\rm Zr}$=(M$_{\rm Ti}$+M$_{\rm Zr}$)/2.}
\begin{tabular}{cccccc}
 superlattice & & &$\overline{Z}_{\rm Zr}$ & $\overline{Z}_{\rm Ti}$ &
              $\omega$(cm$^{-1}$) \\
 \hline
 (001)
 &$\Gamma_{5^\prime}$ &($\perp$)    & 2.32  & 6.76   &  -177  \\
 &$\Gamma_{1^\prime}$ &($\parallel$)& 4.98  & 2.99   &  -178  \\
 &$\Gamma_{5^\prime}$ &($\perp$)    & 3.36  &-1.68   &  -70   \\
 &                    &             &       &        &        \\
 (110)
 &$\Gamma_{1^\prime}$ &($\perp$)    & 2.42  & 7.12   &  -214  \\
 &$\Gamma_{5^\prime}$ &($\parallel$)& 4.39  & 2.80   &  -148  \\
 &$\Gamma_{5^\prime}$ &($\parallel$)& 0.11  & 3.02   &  -76   \\
 &$\Gamma_{1^\prime}$ &($\perp$)    & 2.82  & 0.00   &  -35   \\
 &                    &             &       &        &        \\
 (111)
 &$\Gamma_{15}$       &             & 3.80  & 3.82   &  -110 \\
\end{tabular}
\label{table6}
\end{table}

\onecolumn

\begin{table}
\caption{Frequencies and symmetry labels of the TO and LO 
zone-center eigenmodes. 
For PZT(001) and PZT(110) the LO frequencies are given
for the two directions $\hat{\bf q}_\perp$ and $\hat{\bf q}_\parallel$ 
that are, respectively, perpendicular and parallel to the direction
of compositional modulation.
For PZT(111), the LO frequencies are independent of the direction 
of {\bf q}.
For all modes, the mode effective charges are given in parentheses.} 
\begin{tabular}{c r r r  |c r r r| c r r }
\multicolumn{4}{c|}{(001)} & \multicolumn{4}{c|}{(110)} 
& \multicolumn{3}{c}{(111)} \\
 & & & & & & & & & & \\
   & \multicolumn{1}{c}{TO}           
   & \multicolumn{1}{c} { LO $[\hat{\bf q}_\perp]$} 
   & \multicolumn{1}{c|}{ LO $[\hat{\bf q}_\parallel]$}& 
                    & \multicolumn{1}{c}{TO}   
                    & \multicolumn{1}{c} {LO $[\hat{\bf q}_\perp]$}
                    & \multicolumn{1}{c|}{LO $[\hat{\bf q}_\parallel]$}&
 & \multicolumn{1}{c}{TO} 
 &\multicolumn{1}{c}{LO} \\         
\hline
   $\Gamma_{5^\prime}$ &-185 (9.50) & -82 (0.32) &
  &$\Gamma_{1^\prime}$ &-231 (9.99) & -76 (1.16) &
  &$\Gamma_{15}$       &-110 (7.62) &  95 (1.34) \\
 
   $\Gamma_{1^\prime}$ &-177 (7.92) &            &  94 (1.12)
  &$\Gamma_{5^\prime}$ &-148 (7.19) &            & -85 (0.39)  
  &$\Gamma_{15}$       & 159 (9.27) & 330 (3.32) \\

   $\Gamma_{5^\prime}$ & -73 (1.82) &  86 (0.25) &
  &$\Gamma_{5^\prime}$ & -76 (3.16) &            & 95 (0.72) 
  &$\Gamma_{15}$       & 354 (3.27) & 396 (4.23) \\

   $\Gamma_{5^\prime}$ &  88 (0.59) & 108 (0.95) &
  &$\Gamma_{1^\prime}$ & -37 (2.78) & 128 (2.11) &      
  &$\Gamma_{15}$       & 534 (5.65) & 693 (9.90) \\

   $\Gamma_{1^\prime}$ & 110 (3.53) &            & 225 (1.73)     
  &$\Gamma_{5^\prime}$ &  98 (0.68) &            &  112 (1.00) 
  &                    &            &            \\

   $\Gamma_{5^\prime}$ & 127 (4.03) & 193 (0.22) &            
  &$\Gamma_{5^\prime}$ & 161 (4.00) &            &  194 (0.54)  
  &                    &            &            \\

   $\Gamma_{5^\prime}$ & 194 (0.62) & 373 (5.31) &   
  &$\Gamma_{1^\prime}$ & 161 (4.78) & 370 (5.49) & 
  &                    &            &            \\

   $\Gamma_{1^\prime}$ & 243 (1.83) &            &  306 (4.72)
  &$\Gamma_{5^\prime}$ & 206 (2.96) &            &  320 (3.83)  
  &                    &            &            \\

   $\Gamma_{5^\prime}$ & 440 (1.73) & 466 (3.80) &            
  &$\Gamma_{5^\prime}$ & 378 (4.18) &            &  415 (3.53)  
  &                    &            &            \\

   $\Gamma_{1^\prime}$ & 465 (1.48) &            &  467 (0.72)
  &$\Gamma_{1^\prime}$ & 417 (1.12) & 434 (3.54) & 
  &                    &            &            \\

   $\Gamma_{1^\prime}$ & 545 (6.86) &            &  716 (10.11)
  &$\Gamma_{5^\prime}$ & 532 (5.69) &            &  693 (9.94)   
  &                    &            &            \\

   $\Gamma_{5^\prime}$ & 658 (5.23) & 757 (8.56) &             
  &$\Gamma_{1^\prime}$ & 661 (5.35) & 760 (8.49) &  
  &                    &            &            \\
    \end{tabular}
    \label{table7}
\end{table}

\end{document}